\documentclass[letter]{aa} 

\usepackage{graphicx}
\usepackage{txfonts}
\bibpunct{(}{)}{;}{a}{}{,}
\newcommand{\bz}{\langle B_z \rangle}
\newcommand{\kms}{km\,s$^{-1}$}
\newcommand{\vsini}{$v \sin i$}
\newcommand{\logg}{\log\,g}
\newcommand{\teff}{T_{\rm eff}}

\begin{document} 

\titlerunning{Magnetic He-rich stars}
\authorrunning{J\"arvinen et al.}

   \title{Detection of magnetic fields in He-rich early B-type stars using HARPSpol}

   \author{S.~P.~J\"arvinen
          \inst{1}
          \and
          S.~Hubrig\inst{1}
          \and
          I.~Ilyin\inst{1}
          \and
          M.~Sch\"oller\inst{2}
          \and
          M.~F.~Nieva\inst{3}
          \and
          N.~Przybilla\inst{3}
          \and
          N.~Castro\inst{4}
          }

   \institute{Leibniz-Institut f\"ur Astrophysik Potsdam (AIP),
     An der Sternwarte~16, 14482~Potsdam, Germany\\
              \email{sjarvinen@aip.de}
         \and
         European Southern Observatory, Karl-Schwarzschild-Str.~2,
         85748 Garching, Germany
         \and
         Institut f\"ur Astro- und Teilchenphysik, Universit\"at Innsbruck,
         Technikerstr. 25/8, 6020 Innsbruck, Austria
         \and
         Department of Astronomy, University of Michigan,
         1085 S. University Avenue, Ann Arbor, MI 48109-1107, USA
             }

   \date{Received XXX, 2018; accepted YYY, 2018}

 
  \abstract
   {}
   {We focus on early-B type stars with helium overabundance, for which the
     presence of a magnetic field has not previously been reported.}
   {The measurements were carried out using high-spectral-resolution
     spectropolarimetric observations obtained with the High Accuracy Radial
     velocity Planet Searcher (HARPS) in polarimetric mode, installed at the ESO La
     Silla 3.6m telescope.}
   {For five He-rich stars, the longitudinal magnetic field was detected for
     the first time. For one target, HD\,58260, the presence of a longitudinal
     magnetic field of the order of 1.8\,kG has already been reported in the
     literature, but the magnetic field has remained constant over tens of
     years. Our measurement carried out using the polarimetric spectra
     obtained in 2015 March indicates a slight decrease of the longitudinal
     magnetic field strength compared to measurements reported in previous
     works. A search for periodic modulation in available photometric data
     allowed us to confidently establish a period of 2.64119$\pm$0.00420\,d in
     archival ASAS3 data for CPD$-$27\degr 1791. No period could be determined
     for the other five stars.}
   {The obtained results support the scenario that all He-rich stars are
     detectably magnetic and form an extension of the Ap star phenomenon to
     higher temperatures.}


   \maketitle
%

\section{Introduction}\label{sec:intro}

During the last years an increasing number of massive stars has been
investigated for magnetic fields in the framework of the Magnetism in 
Massive Stars (MiMeS) and B-fields in OB stars (BOB) 
surveys 
\citep{mimes, BOB}. 
Direct magnetic-field measurements are of great importance to properly 
understand the potential effects of magnetic fields on the evolution of 
massive stars, including the impact on angular momentum and on stellar wind 
properties. While the BOB  survey mostly 
concentrated on normal main sequence OB stars, the MiMeS  survey consisted of a survey component and a targeted 
component to characterise a sample of known magnetic stars. Since previously 
detected magnetic O and early B-type stars on average appeared to have 
rotation velocities significantly lower than the rest of the population, to 
enhance the probability of detecting magnetic fields, the majority of the 
stars targeted by BOB were relatively slowly rotating, with \vsini{} values below 
60\,\kms 
\citep{BOB}.
However, according to
\citet{Markus2017},
the results of the search for the presence of a magnetic field in these stars
did not result in higher yields, with a magnetic field detection rate of
5$\pm$5\%. 

While these results indicate that the presence of a magnetic field in normal,
slowly rotating O and early B-type stars is not common, magnetic early B-type
He-rich stars constitute about 10\% of all main sequence early-B type stars.
This group contains the most massive chemically peculiar stars with spectral
types around B2 and exhibits helium and silicon surface spots
\citep[e.g.][]{landstreet, borra, bohlender, Hubrig2017, Castro2017}.
The distribution of these spots is non-uniform and non-symmetric with respect
to the rotation axis. Also, lines belonging to CNO and iron peak elements
frequently show variable line profiles over the rotation periods, but the
distribution of these elements is poorly documented in the literature due to
the relative weakness of their lines compared to He and Si lines.  

From previous studies of He-rich stars, we know that inhomogeneous chemical
abundance distributions in early-B type stars are only observed on the surface
of magnetic chemically peculiar stars with large-scale organised magnetic
fields. Among such
stars, five stars, 
\object{CPD$-$27$^{\circ}$1791}, 
\object{HD\,60344}, 
\object{HD\,149257},
\object{CPD$-$69$^{\circ}$2698}, and 
\object{HD\,168785}, have previously been reported to exhibit a surface 
helium overabundance 
\citep[e.g.][]{MacConnell1970, Garrison1977}.
As no definite magnetic fields were reported in previous studies of these
stars, they were included in the BOB target list. Obviously, these He-rich
stars are excellent candidates to confirm the positive correlation between the
presence of a magnetic field and helium enrichment in the star's atmosphere.
The sixth star in our sample, 
\object{HD\,58260}, 
was shown to be magnetic by 
\citet{borra},
but exhibited a surprisingly constant magnetic field over nine years
\citep{bohlender1987}.
A recent study of this star by 
\citet{shultz}
suggested that its magnetic field shows no variability over a timescale of
about 35 years. In the following, we report on our results of the magnetic
field measurements in all six stars.

\section{Observations and magnetic field analysis}
\label{sec:obs}

\begin{table}
\centering
\caption{
  Logbook of the observations. The columns give the name of the star,
  the visual magnitude, the heliocentric Julian date (HJD) for the middle of
  the exposures, and the S/N of the spectra.
}
\label{tab:log}
\begin{tabular}{lccc}
\hline
\hline
Star &
$m_{\rm V}$ &
HJD &
S/N \\
& & 2\,400\,000+ & \\
\hline
CPD$-$27\degr 1791 & 9.3  & 57095.6063 & 162 \\
HD\,58260          & 6.7  & 57091.6462 & 329 \\
HD\,60344          & 9.1  & 57093.6454 & 258 \\
HD\,149257         & 9.2  & 57095.7374 & 234 \\
CPD$-$69\degr 2698 & 9.3  & 57178.6338 & 154 \\
HD\,168785         & 8.5  & 57095.8469 & 265 \\
HD\,168785         & 8.5  & 57316.5335 & 209 \\
\hline
\end{tabular}
\end{table}

The observations of the He-rich early B-type stars used here were obtained on 2015 
March 10, 12, and 14, 2015 June 5, and 2015 October 20 using the High 
Accuracy Radial velocity Planet Searcher in polarimetric mode
\citep[HARPSpol;][]{harps}
installed at the ESO La Silla 3.6\,m telescope. All spectra have a resolving 
power of about $R = 110\,000$ and cover the spectral range 3780--6910\,\AA{}, 
with a small gap between 5259\,\AA{} and 5337\,\AA{}. The reduction and 
calibration of these spectra was performed using the HARPS data-reduction 
software available on La~Silla. The normalization of the spectra to the 
continuum level is described in detail by 
\citet{Hubrig2013}. 
A summary of the HARPSpol observations is given in Table~\ref{tab:log}.

To study the presence of a magnetic field in the Bp stars, we employed the
least-squares deconvolution (LSD) technique
\citep{Donati1997}. 
This technique combines line profiles (using the assumption that line
formation is similar in all lines) centred at the position of the individual
lines given in the line mask and scaled according to the line strength and
the sensitivity to a magnetic field (i.e.\ to the Land{\'e} factor). The
resulting average profiles (Stokes $I$, Stokes $V$, and diagnostic null
spectrum $N$) obtained by combining several lines, yield an increase in the
signal-to-noise ratio (S/N) as the square root of the number of the lines used.
For the studied stars, the line masks corresponding to $\teff{}$ and $\logg{}$
determined by 
\citet[][(see Sect.~\ref{sec:meas})]{Zboril1999}, 
were constructed  using the Vienna Atomic Line Database 
\citep[VALD;][]{Kupka2011, VALD3}. 
Only lines actually visible in the observed spectra have been used to
calculate the LSD profiles. Lines blended with hydrogen lines and lines in
spectral regions contaminated by telluric features were excluded from the line
list. The mean longitudinal magnetic field was evaluated by computing the
first-order moment of the Stokes $V$ profile according to
\citet[][]{Mathys1989}.

\begin{figure}
 \centering 
        \includegraphics[width=1.0\columnwidth]{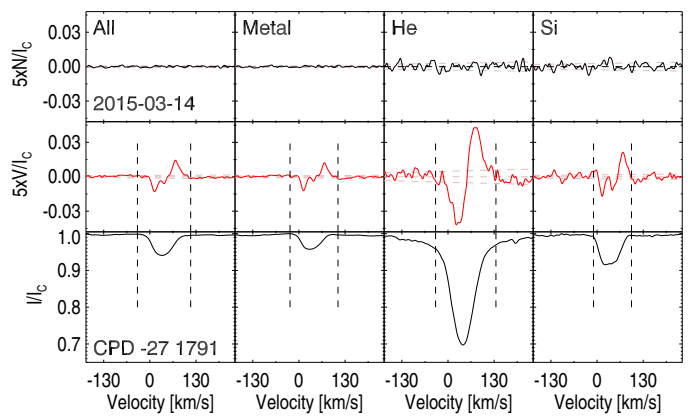}
        \caption{
          LSD Stokes $I$, $V$, and diagnostic null $N$ profiles (from bottom
          to top) calculated for CPD$-$27$^{\circ}$1791 using a line mask
          containing all spectral lines, a line mask with only metal lines, a
          line mask containing exclusively He lines and a line mask containing
          exclusively Si lines (from left to right). Vertical dashed lines
          indicate the integration ranges for the determination of $\bz$.
         }
   \label{fig:IVN_cpd271791}
\end{figure}

\begin{figure}
 \centering 
        \includegraphics[width=1.0\columnwidth]{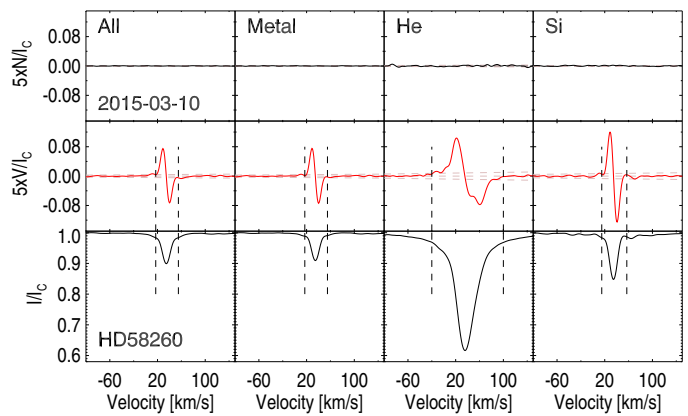}
        \caption{
          As in Fig.~\ref{fig:IVN_cpd271791}, but for HD\,58260.
         }
   \label{fig:IVN_HD58260}
\end{figure}

\begin{figure}
 \centering 
        \includegraphics[width=1.0\columnwidth]{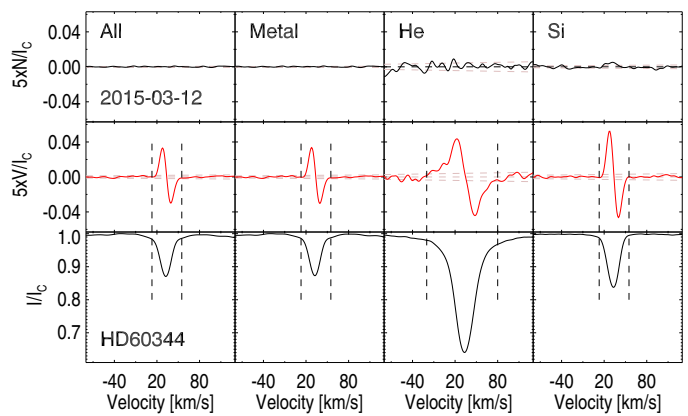}
        \caption{
          As in Fig.~\ref{fig:IVN_cpd271791}, but for HD60344.
         }
   \label{fig:IVN_HD60344}
\end{figure}

\begin{figure}
 \centering 
        \includegraphics[width=1.0\columnwidth]{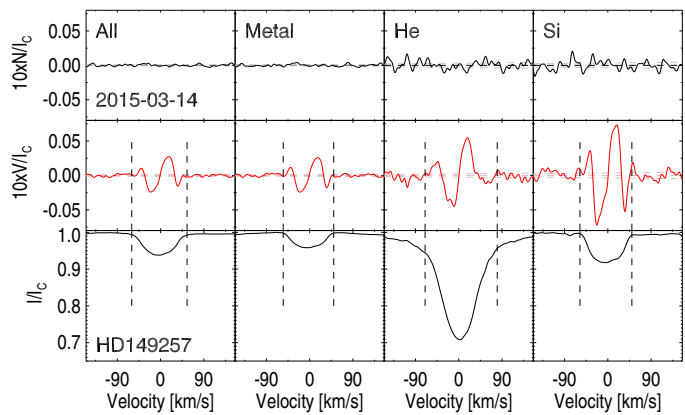}
        \caption{
          As in Fig.~\ref{fig:IVN_cpd271791}, but for HD\,149257.
         }
   \label{fig:IVN_HD149257}
\end{figure}

\begin{figure}
 \centering 
        \includegraphics[width=1.0\columnwidth]{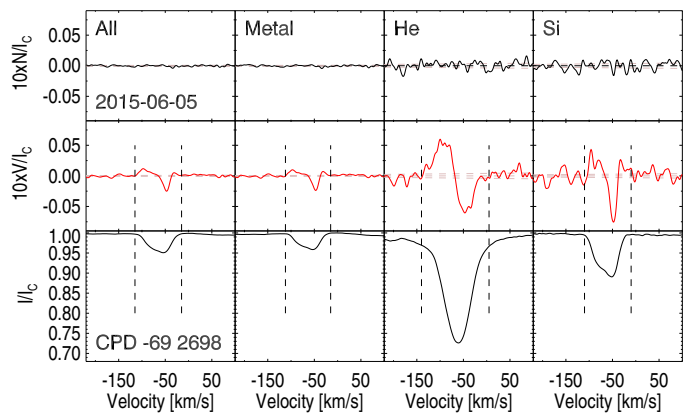}
        \caption{
          As in Fig.~\ref{fig:IVN_cpd271791}, but for CPD$-$69$^{\circ}$2698.
         }
   \label{fig:IVN_cpd692698}
\end{figure}

\begin{figure}
 \centering 
        \includegraphics[width=1.0\columnwidth]{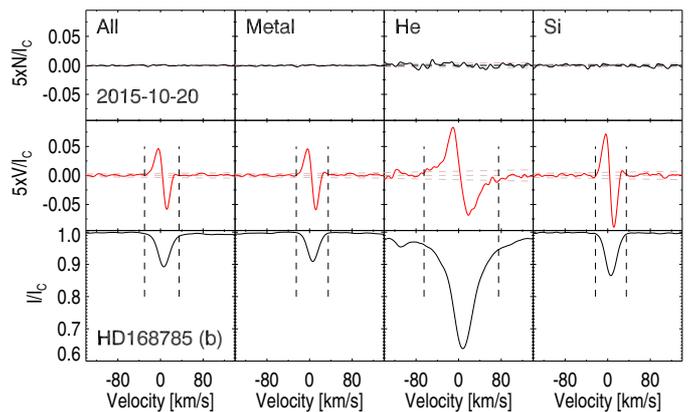}
        \includegraphics[width=1.0\columnwidth]{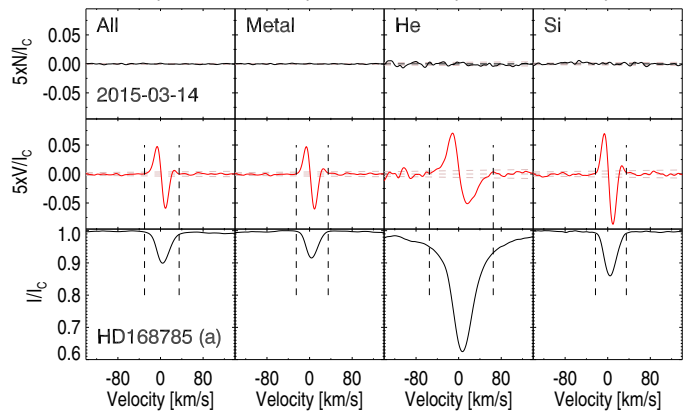}
        \caption{
          As in Fig.~\ref{fig:IVN_cpd271791}, but for HD168785.
         }
   \label{fig:IVN_HD168785_b}
\end{figure}

\begin{table*}
\centering
\caption{
  Magnetic field measurements for the He-rich stars using four different
  line masks. The measurement for HD\,168785 denoted by $^a$ was taken on
  2015 March 14 while the one denoted by $^b$ was taken on 2015 October 10.
}
\label{tab:mfield}
\begin{tabular}{ l r@{$\pm$}l r@{$\pm$}l r@{$\pm$}l r@{$\pm$}l}
\hline
\hline
Star & 
\multicolumn{2}{c}{$\left<B_{\rm z}\right>_{\rm All}$} &
\multicolumn{2}{c}{$\left<B_{\rm z}\right>_{\rm metal}$} &
\multicolumn{2}{c}{$\left<B_{\rm z}\right>_{\rm He}$} &
\multicolumn{2}{c}{$\left<B_{\rm z}\right>_{\rm Si}$}\\      
&
\multicolumn{2}{c}{[G]} &
\multicolumn{2}{c}{[G]} &
\multicolumn{2}{c}{[G]} &
\multicolumn{2}{c}{[G]} \\
\hline
CPD$-$27\degr 1791 & $-$1125 & 29 & $-$1282 & 37 & $-$933 & 27 & $-$695 & 29 \\
HD\,58260          & 1488    & 37 & 1537    & 48 & 1426   & 29 & 1446   & 31 \\
HD\,60344          & 370     & 12 & 377     & 12 & 667    & 30 & 403    & 13 \\
HD\,149257         & $-$161  & 17 & $-$189  & 27 & $-$70  & 18 & $-$256 & 43 \\
CPD$-$69\degr 2698 & 466     & 43 & 414     & 54 & 868    & 44 & 402    & 87 \\
HD\,168785$^a$     & 863     & 11 & 955     & 15 & 712    & 19 & 675    & 10 \\
HD\,168785$^b$     & 1040    & 12 & 1153    & 17 & 1145   & 24 & 915    & 13 \\
\hline
\end{tabular}
\end{table*}


\section{Longitudinal magnetic field measurements}
\label{sec:meas}

As already mentioned in Sect.~\ref{sec:intro}, in He-rich stars, the
magnetic field and the large-scale distribution of certain elements, in 
particular helium and silicon, is non-symmetric with respect to the rotation 
axis. The majority of studies of He-rich stars have revealed a kind of
symmetry between the topology of the magnetic field and the element
distribution
\citep[see e.g.][]{Wade, Bohlender2010, Yakunin2015, Hubrig2017}.
Therefore, the correspondence between the magnetic field structure and the 
location of chemical spots can be roughly probed by measuring the 
longitudinal magnetic field using spectral lines of inhomogeneously
distributed elements separately. As an example, the main concentration of He
is frequently observed in the vicinity of the positive magnetic pole and
Si-deficient spots around both poles. High-concentration Si spots are
usually located close to the magnetic equator. We note, however, that given 
the availability of only single observing epochs for each target, the 
suggestions on the element locations in respect to the magnetic field 
topology should be considered very tentative. 

To take the effect of the inhomogeneous surface element distribution into 
account, we decided to use four different line masks for the measurements,
one including all photospheric lines apart from the hydrogen lines, one mask
with all metal lines visible in the spectra, one mask containing exclusively
He lines, and one mask containing exclusively the Si lines. The magnetic field
measurements for the He-rich stars using these four different line masks are
presented in Table~\ref{tab:mfield} and the resulting LSD Stokes $I$, $V$, and
diagnostic $N$ profiles are illustrated in
Figs.~\ref{fig:IVN_cpd271791} -- \ref{fig:IVN_HD168785_b}. For all cases, the
false alarm probability (FAP) is less than $10^{-5}$, confirming definite
detections, and the null spectra appear flat, indicating the absence of
spurious polarization. In the following, we briefly discuss the measurements
of the longitudinal magnetic field using the different line masks for each
star separately.

\subsection{CPD$-$27$^{\circ}$1791}

This star was identified as helium-rich by 
\citet{Garrison1977}.  
The longitudinal magnetic field in CPD$-$27$^{\circ}$1791 is measured for the
first time.  As is shown in Fig.~\ref{fig:IVN_cpd271791}, apart from the
clear Zeeman feature calculated for the line mask with the He lines, all other
Zeeman features calculated for line masks containing metal lines present a
distinct substructure likely related to an inhomogeneous surface element
distribution. In agreement with this suggestion, the LSD Stokes $I$ profiles
calculated for these line masks appear asymmetric (best visible in the Stokes
$I$ profile calculated for the Si lines), indicating the presence of surface
chemical spots. The lowest value for the longitudinal magnetic field,
$\bz=-695\pm29$\,G, is measured using the line mask containing the Si lines 
suggesting that we probably observe this star close to the magnetic pole and 
that the Si lines may form in a spot at some distance from the magnetic pole.
The atmospheric parameters $\teff=22\,700$\,K and $\logg=4.53$, typical for a
spectral type B1.5--B2, were determined by 
\citet{Zboril1999},
who used Kurucz models 
\citep{Kurucz1992}.

\subsection{HD\,58260}

\begin{figure}
 \centering 
        \includegraphics[width=1.\columnwidth]{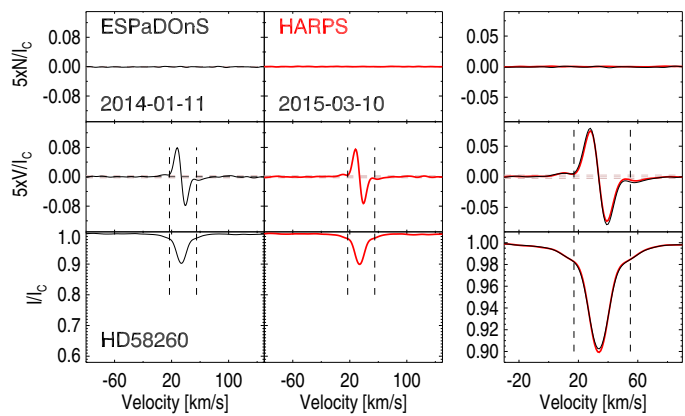}
        \caption{
          The LSD Stokes~$I$, $V$, and diagnostic $N$ profiles
          calculated for HD\,58260 using the ESPaDOnS spectrum (black line,
          left panel) and the HARPS spectrum (red line, middle panel). A
          zoom-in to the overplotted profiles is presented in the panel on the
          right. 
         }
   \label{fig:dif}
\end{figure}

This star was identified as helium-rich by
\citet{Garrison1977}.  
According to 
\citet{bohlender1987},
the magnetic field of this star has been constant at 2300\,G between 1977
December and 1986 March. Using these historical measurements along with nine
more recent Echelle SpectroPolarimetric Device for the Observation of Stars
\citep[ESPaDOnS;][]{espadons, espperf}
measurements, 
\citet{shultz} 
reported a $\bz$ of the order of 1.8\,kG measured in spectra obtained in
2013--2014. The authors were unable to determine a variability period for
this star due to the negligible variation of $\bz$. All available measurements
were within 1$\sigma$ of the mean ESPaDOnS $\bz$, suggesting a non-variability
of the longitudinal magnetic field over a timescale of about 35 years. Our
measurements show $\bz$ values in the range 1430--1540\,G, and are weaker than the values reported by 
\citet{shultz} by
$\sim$300\,G.
To check the robustness of our result indicating a 300\,G difference 
between our measurements and the measurements of
\citet{shultz},
we downloaded the ESPaDOnS spectrum of HD\,58260 from the
PolarBase\footnote{http://polarbase.irap.omp.eu}
\citep{Petit2014, Donati1997}.
This spectrum was obtained on 2014 January 11 and is presented in Fig.~1 in 
the paper of
\citet{shultz}.
Using the same line list as that used in our HARPS observations we obtain
$\left<B_{\rm z}\right>_{\rm metal}=1882\pm40$\,G. This value is in agreement with
the value $\left<B_{\rm z}\right>_{\rm metal}=1820\pm30$\,G published by Shultz et
al. In Fig.~\ref{fig:dif} we present the comparison between the LSD Stokes $I$
and $V$ profiles calculated for HD\,58260 using the HARPS and ESPaDOnS spectra.
The strongest longitudinal magnetic field is detected using the line mask with
the metal lines suggesting that we probably observe this star at some distance
from the magnetic pole. The atmospheric parameters $\teff=19\,000$\,K and
$\logg=4.02$, typical for a spectral type B2.5, were determined by
\citet{Zboril1999}.

\subsection{HD\,60344}

The longitudinal magnetic field in this star is measured for the first time.
The helium-rich nature of this star was for the first time mentioned by 
\citet{MacConnell1970}.
A longitudinal magnetic field was not detected in the previous studies by 
\citet{borra}
and 
\citet{bohlender1987}.
Our observations clearly show evidence of the presence of a magnetic field of
the order of a few hundred Gauss. The strongest longitudinal magnetic field,
$\bz=667\pm30$\,G, is measured using the He lines, indicating that the He
lines may form in a spot close to the positive magnetic pole. The atmospheric
parameters $\teff=21\,700$\,K and $\logg=4.48$, typical for a spectral type
B2, were determined by
\citet{Zboril1999}. 

\subsection{HD\,149257}

The longitudinal magnetic field of this star is measured for the first time.
The helium-rich nature of this star was for the first time mentioned by 
\citet{Garrison1977}.
\citet{Wiegert1998} 
discovered significant variations in the hydrogen and helium lines. As is shown
in Fig.~\ref{fig:IVN_HD149257}, the Zeeman features calculated for all masks
exhibit profiles reminiscent of a typical crossover profile and correspond to 
the lowest longitudinal magnetic fields measured in our star sample. Such 
profile shapes are usually observed when both the negative and the positive 
magnetic pole become visible at the time of observation, that is,\ we observe 
HD\,149257 at this epoch close to the magnetic equator. The presence of a 
distinct substructure in the LSD Stokes $I$ profile obtained for the mask 
with the silicon lines is most likely related to an inhomogeneous surface Si 
distribution. The strongest longitudinal magnetic field, $\bz=-256\pm43$\,G, 
is measured using the mask with the Si lines supporting the assumption that 
Si spots are frequently distributed close to the magnetic equator.
\citet{bagnulo2006}
observed HD\,149257 in 2004 using the FOcal Reducer low dispersion Spectrograph 
\citep[FORS\,2;][]{1998Messenger}
mounted on the 8\,m Antu telescope of the VLT and did not detect a magnetic
field. The atmospheric parameters $\teff=24\,900$\,K and $\logg=4.19$, typical
for a spectral type B1.5, were determined by 
\citet{Zboril1999}. 

\subsection{CPD$-$69$^{\circ}$2698}

The longitudinal magnetic field in this star is measured for the first time.
The helium-rich nature of this star was for the first time mentioned by 
\citet{MacConnell1970}.  
Apart from the very clear Zeeman feature calculated for the line mask with the
He lines, with a shape characteristic for a longitudinal magnetic field of
positive polarity, all other Zeeman features calculated for the line masks
containing metal lines exhibit a distinct substructure, probably related to
an inhomogeneous surface element distribution. In agreement with this
suggestion, the LSD Stokes $I$ profiles calculated for these line masks appear
asymmetric. The longitudinal magnetic field measured using the He lines,
$\bz=868\pm44$\,G, is stronger by almost a factor of two than the longitudinal
magnetic field measured using the metal lines. This suggests that the He lines
may form in a spot that is closer to the magnetic pole, while the metal lines
form at some distance from the magnetic pole. The atmospheric parameters
$\teff=25\,300$\,K and $\logg=3.90$, typical for a spectral type B1, were
determined by
\citet{Zboril1999}. 

\subsection{HD\,168785}

The helium-rich nature of this star was first mentioned by 
\citet{MacConnell1970}. 
\citet{Wiegert1998}
discovered significant variations in the hydrogen and helium lines. The
longitudinal magnetic field in this star is measured for the first time. This
star was observed twice, on 2015 March 14 and on 2015 October 20. The
longitudinal magnetic field measured in October is
stronger than the longitudinal magnetic field measured in March by a few hundred Gauss. For both
observations the lowest longitudinal magnetic field strength,
$\bz=675\pm10$\,G and $\bz=915\pm13$\,G, is measured using the mask with the Si
lines suggesting that a Si abundance spot is probably clearly visible at these 
observing epochs and we observe this star at some distance from the magnetic pole.
The atmospheric parameters $\teff=23\,400$\,K and $\logg=4.11$, typical for a 
spectral type B1.5, were determined by
\citet{Zboril1999}. 


\section{Discussion}
\label{sec:disc}

\begin{table*}
\centering
\caption{
  Summary of photometric measurements and \vsini{} estimates. For each star in
  column~1, we list the interval and number of measurements, for ASAS3 in
  columns~2 and 3, and for \emph{Hipparcos} in columns~4 and 5. Column~6
  presents the  \vsini{} values obtained by
  \citet{Zboril1999}
  and column~7 has the values measured on our HARPS spectra.
}
\label{tab:sum}
\begin{tabular}{llclccc}
\hline
\hline
Star &
ASAS3 &
No. & 
\emph{Hipparcos} &
No. &
$v\sin i_{\rm{ZN}}$ &
$v\sin i^{*}$ \\
&
Interval &
meas. &
Interval &
meas. &
[\kms] &
[\kms] \\
\hline
CPD$-$27\degr 1791 & 2000 Nov/2009 Dec & 997 & 1990 Jan/1993 Mar & 178 & 45$\pm$4 & 37$\pm$3 \\
HD\,58260$^{**}$   & 2000 Nov/2009 Dec & 781 & 1989 Nov/1993 Mar & 124 & 45$\pm$6 & 9$\pm$1 \\
HD\,60344          & 2000 Nov/2009 Dec & 617 & 1990 Mar/1993 Mar & 107 & 55$\pm$6 & 10$\pm$1 \\
HD\,149257         & 2001 Jan/2009 Oct & 722 & & & 40$\pm$4 & 48$\pm$2 \\
CPD$-$69\degr 2698 & 2001 Jan/2009 Nov & 653 & 1989 Dec/1993 Feb & 265 & 30$\pm$3 & 26$\pm$2 \\
HD\,168785         & 2001 Feb/2009 Nov & 996 & & & 14$\pm$2 & 14$\pm$1 \\
\hline
\end{tabular}
\tablefoot{$^{**}$ \citet{shultz} give $v \sin i=3\pm2$\kms.}
\tablebib{ZN = \citet{Zboril1999},  $^{*}$ = this work}
\end{table*}

Our spectropolarimetric observations with HARPSpol allowed
us to detect for the first time the presence of relatively strong longitudinal
magnetic fields in five He-rich stars. For the sixth He-rich star, HD\,58260,
we found an indication of a magnetic field decrease compared to previous
measurements.  

The measurements using different line masks belonging to different elements
indicate the presence of chemical spots on the surface of all stars in our
sample. Given the presence of clearly detected longitudinal magnetic fields in
our targets, the determination of  their magnetic periods and magnetic field
geometries should be easily feasible and can be carried out in follow-up
observations. Periodic magnetic variations in such chemically peculiar stars
are related to the rotation periods and are generally described by the oblique
rotator model 
\citep{Stibbs1950}
with a magnetic field having an axis of symmetry at an angle to the rotation
axis, usually called the obliquity angle $\beta$.

\begin{figure}
 \centering 
        \includegraphics[width=1.0\columnwidth]{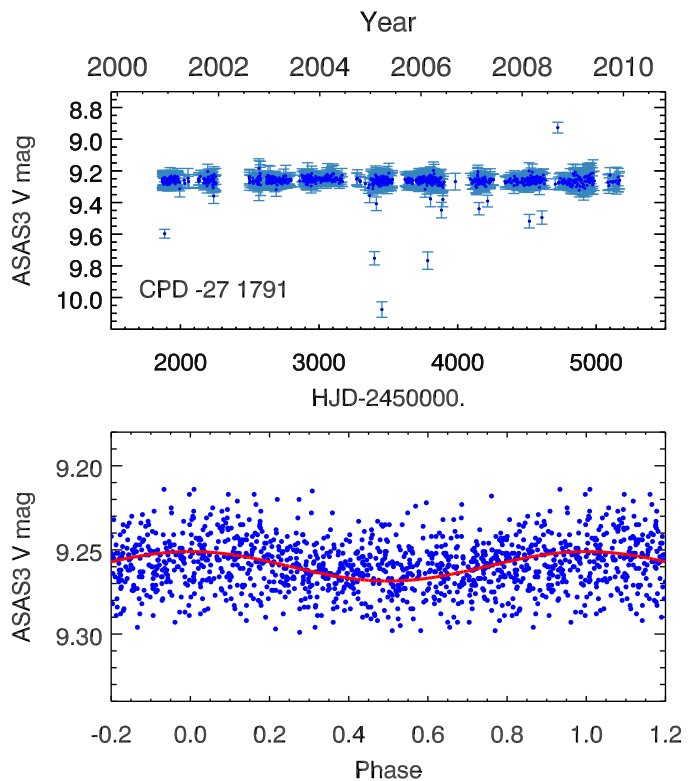}
           \caption{
         Rotational period of CPD$-$27\degr 1791. Top: ASAS3 light curve.
Bottom: ASAS3 photometry phased with $P_{\rm rot}=2.64119$\,d.
         }
   \label{fig:phot}
\end{figure}

Since the rotation periods for our sample stars are unknown, we searched
for periodic modulations using photometric data. He-rich stars and
chemically peculiar stars in general usually exhibit photometric variability
due to surface chemical abundance spots. Therefore, rotational periods can
be determined from photometry. We have used archival data from the ASAS3
survey\footnote{http://www.astrouw.edu.pl/asas/}
\citep{asas},
covering the time interval from 2000 to 2009 and from the \emph{Hipparcos}
survey
\citep{Perryman1997,vanLeeuwen2007}
covering the time interval from 1989 to 1993. The photometry time intervals
and the number of measurements are summarised in Table~\ref{tab:sum} in
columns~2--5. After removing some apparent outliers, we carried out a period
search using a non-linear least-squares fit to multiple harmonics using
the Levenberg--Marquardt method
\citep{Press}.
To detect the most probable period, we calculated the frequency spectrum,
and for each trial frequency we performed a statistical F-test of the null
hypothesis for the absence of periodicity
\citep{seber}.
The resulting F-statistics can be thought of as the total sum, including
covariances of the ratio of harmonic amplitudes to their standard deviations,
i.e.\ an S/N
\citep[e.g.][]{Hubrig2017}.
Using ASAS3 observations, a single significant peak has been detected in the
frequency spectrum of CPD$-$27\degr 1791 corresponding to
$P_{\rm rot}=2.64119\pm0.00420$\,d at a high confidence level with an FAP value
of $2.3\times 10^{-9}$. No periodicity was detected in ASAS3 and in
\emph{Hipparcos} photometry for the other five stars. In Fig.~\ref{fig:phot}
we present in the upper panel the light curve collected by ASAS3 over nine
years and in the lower panel the ASAS3 photometry phased with the rotation
period of 2.64119\,d.

The absence of $\left<B_{\rm z}\right>$ phase curves (i.e.\ the
dependence of the magnetic field strength on the rotation phase) for our
sample stars does not allow us to conclude on the real field strength, which
can be much higher, or the field topology. Since the determination of
abundances and \vsini{} values depends on the chemical spot distribution and
magnetic line intensification (which is different at different rotational
phases), spectropolarimetric time series obtained over the rotation periods
are necessary for an in-depth analysis of the atmospheres, that is,\ an
exploration of the magnetic field strength and the surface abundance
distribution would require Zeeman Doppler Imaging of the chemical abundance
pattern. 

Since HARPS spectra were obtained at an excellent resolving power
of 110\,000, we estimated projected rotational velocities for each target by
fitting Gaussian profiles with full widths at half-maximum (FWHM) to the
unblended \ion{Ne}{i} 6402 line and compared them with literature values.
Apart from the values determined for stars HD\,58260 and HD\,60344, our
measurements presented in column~7 of Table~\ref{tab:sum} appear to be in
good agreement with the determinations of
\citet{Zboril1999}
presented in column~6, who used lower resolution
spectra with $R=30\,000$ obtained with the CES spectrograph at the ESO-CAT
telescope on La~Silla in Chile to study CNO abundances in the atmospheres of 
our sample stars.

The obtained results show that magnetic fields can be detected in all He-rich
stars and that, in agreement with the suggestion of 
\citet{Osmer1974},
these stars form an extension of the Ap star phenomenon to higher 
temperatures. Similar to Ap stars, they are spectrum variables and also show 
variable photometrical light curves 
\citep[e.g.][]{Pedersen1977}.
However, while the evolution of the magnetic field strength and field 
geometry, including the evolution of the dipole obliquity angle $\beta$ in Ap 
and late-B-type stars across the main sequence, has already been the subject of 
several careful studies using representative stellar samples 
\citep[e.g.][]{Hubrig2000, Hubrig2007},
no comparable study exists for a representative sample of early B-type He-rich
stars with well-defined rotation periods and magnetic field geometries. As we 
showed in the previous section, the longitudinal magnetic field measurements 
in He-rich stars are strongly affected by the presence of chemical spots,
making the determination of the obliquity angle $\beta$ difficult. 
Furthermore, the magnetic field geometry in these stars frequently shows 
contributions from non-dipolar multipoles. In particular, the distribution of the 
obliquity angle $\beta$ is essential to understand the physical processes 
taking place in these stars and the origin of their magnetic fields. 
As discussed by 
\citet{Moss1986} 
and later by 
\citet{Hubrig2007},
randomness in the $\beta$ distribution may be regarded as an argument in
favour of the fossil field, since the star-to-star variations in obliquity of
the magnetic field axes can plausibly be interpreted as reflecting differences
in the intrinsic magnetic conditions at different formation sites. An
important issue of the fossil field theory discussed in the past was the
survival of the magnetic field over the star's lifetime, as it was difficult
to find stable field configurations
\citep[e.g.][]{Mestel1984, Moss1986}. 
However, more recently,
\citet{2004Nature}
used three-dimensional numerical MHD simulation and showed that stable
magnetic field configurations can develop through evolution from arbitrary,
random initial magnetic fields. The magnetic field permeating the interstellar
medium is amplified during star formation and may naturally relax into a
large-scale, mostly poloidal field emerging at the surface. The work of
\citet{2004Nature}
(see also the work of
\citet{Braithwaite2008,Braithwaite2009,Duezetal})
resulted in fundamental revision of our understanding  of the behaviour of
magnetic fields in stellar radiative zones. On the other hand, the magnetic
field excited by a dynamo mechanism is expected to be either symmetric or
antisymmetric in regard to the equatorial plane 
\citep[e.g.][]{Krause1976}.


\begin{acknowledgements}
  We thank the referee, G.\ Wade, for useful comments. Based on observations
  made with ESO Telescopes at the La Silla Paranal Observatory under programme
  ID 191.D-0255. This work has made use of the VALD database, operated at
  Uppsala University, the Institute of Astronomy RAS in Moscow, and the
  University of Vienna.
\end{acknowledgements}

%
   \bibliographystyle{aa} 
   \bibliography{Bp} 
%

\end{document}